\documentstyle[prb,aps,epsfig]{revtex}

\begin{document}
\draft
\title{Quantitative imaging of dielectric permittivity and tunability with a near-field scanning microwave microscope}

\author{D. E. Steinhauer, C. P. Vlahacos, F. C. Wellstood, and Steven M. Anlage$^{a)}$\footnotetext{$^{a)}$
Electronic mail: anlage@squid.umd.edu. Color versions of the figures in
this paper can be found at http://www.csr.umd.edu/research/hifreq/micr\_microscopy.html.}}
\address{Center for Superconductivity Research, Department of Physics, University of Maryland, College Park, MD 20742-4111 }
\author{C. Canedy and R. Ramesh}
\address{Department of Materials and Nuclear Engineering, University of Maryland, College Park, MD 20742-2115}
\author{A. Stanishevsky and J. Melngailis} 
\address{Department of Electrical Engineering, University of Maryland, College Park, MD 20742-3285}

\maketitle

\begin{abstract}
We describe the use of a near-field scanning microwave microscope to image the permittivity and tunability of bulk and thin film dielectric samples on a length scale of about 1 $\mu$m.  The microscope is sensitive to the linear permittivity, as well as to nonlinear dielectric terms, which can be measured as a function of an applied electric field.  We introduce a versatile finite element model for the system, which allows quantitative results to be obtained.   We demonstrate use of the microscope at 7.2 GHz with a 370 nm thick Ba$_{0.6}$Sr$_{0.4}$TiO$_3$ thin film on a LaAlO$_3$ substrate.  This technique is nondestructive and has broadband (0.1-50 GHz) capability.  The sensitivity of the microscope to changes in permittivity is $\Delta\epsilon_{r} = 2$ at $\epsilon_{r} = 500$, while the nonlinear dielectric tunability sensitivity is $\Delta \epsilon _{113} = 10^{-3}\ (kV/cm)^{-1}$.  
\end{abstract}


\section{Introduction}

Dielectric thin film research has become increasingly important as the demand grows for smaller, faster, and more reliable electronics.  For example, high permittivity thin films are under study in order to fabricate smaller capacitors while minimizing leakage.  Low permittivity materials are being sought to allow smaller scale circuits while minimizing undesirable stray capacitance between wires.  Nonlinear dielectrics, which have a dielectric permittivity which is a function of electric field, are being used in tunable devices, particularly at microwave frequencies.  Finally, ferroelectric thin films are a solution for large-scale, non-volatile memories.

All of these dielectric thin film technologies demand high-quality, homogeneous films.  However, thin film properties and homogeneity can be difficult to measure.  Consequently, it is important to have a set of reliable techniques for evaluating thin film properties, such as permittivity and tunability.  A number of different techniques are available.  Thin film capacitors which are sensitive to the normal\cite{Miranda,Basceri} and in-plane\cite{Chang} components of the permittivity tensor can be used at low frequencies ($\lesssim$ 1 MHz).  Microwave measurements of permittivity have been made using transmission measurements through a microstrip structure\cite{Carrol}, and reflection measurements from a Corbino probe\cite{Jiang,Booth} and similar geometries.\cite{Ma}  However, these techniques average over large areas, and involve depositing thin film electrodes, which itself can alter the properties of the sample.  Dielectric\cite{Findikoglu} and open\cite{Harrington} resonators have been used as well, but suffer from low spatial resolution.  Cavity perturbation\cite{Zuccaro} and waveguide transmission\cite{BelhadjTahar} techniques have also been used, but have the disadvantage of measuring the entire sample.  More recently, near-field microscopy techniques\cite{Cho,Gao,Gao2,Steinhauer1,Steinhauer2,AnlageASC,VlahacosDielectric,Steinhauer3} have allowed quantitative measurements with spatial resolutions much less than the wavelength.  These techniques use a resonator which is coupled to a localized region of the sample through a small probe, and have the advantage of being non-destructive.  However, it is still difficult to arrive at quantitative results and maintain high spatial resolution.

Previously, we introduced our solution, a near-field scanning microwave microscope (NSMM), and its application to the low-resolution (100--500 $\mu$m) quantitative imaging of sheet resistance\cite{Steinhauer1,Steinhauer2} and dielectric permittivity.\cite{VlahacosDielectric}  In this paper, we present a method for using the NSMM for high-resolution ($\sim 1\ \mu$m), quantitative imaging of thin film permittivity and tunability.\cite{Steinhauer3}

\section{Description of the Microscope}

Our NSMM systems (see Fig.\ \ref{schematic}) consist of a coaxial transmission line resonator which is terminated at one end with an open-ended coaxial probe, and at the other end is coupled to a microwave source with a coupling capacitor.  The probe, which has a sharp-tipped center conductor extending beyond the outer conductor, is held fixed while the sample is raster scanned beneath the probe.  A sample holder gently presses the sample against the probe tip with a force (about 50 $\mu$N \cite{force_ref}) which is small enough that the sample is not scratched.\cite{Steinhauer3}  Because the rf fields are concentrated at the probe tip, the resonant frequency and quality factor Q of the resonator are a function of the sample properties near the probe tip.  A feedback circuit\cite{Steinhauer1,Steinhauer2} keeps the microwave source locked onto a selected resonant frequency of the microscope resonator, and outputs the frequency shift ($\Delta f$) and Q of the resonator.  This data is stored by a computer, which also controls the scanning of the sample beneath the probe.  We have shown previously that the spatial resolution of the microscope in this mode of operation is about 1 $\mu$m.\cite{AnlageASC,Steinhauer3}

Our probes\cite{AnlageASC,Steinhauer3,patent} are constructed using a short section of semi-rigid coaxial cable, where we have replaced the center conductor with a capillary tube of the same outer diameter.  The probe tip is a commercial scanning tunneling microscope (STM) conical tungsten tip\cite{stm_tip} which is inserted into the capillary tube, and held in place by friction.  

We apply a local dc electric field to thin film samples with a bias tee\cite{bias_tee} in the microscope resonator (see inset to Fig.\ \ref{schematic}), allowing us to make dielectric tunability measurements.  A counterelectrode beneath the thin film acts as a ground plane.  

\section{Physical Model for the System}
\subsection{Calculating the Electric Field Near the Probe Tip}

In order to arrive at quantitative results, we developed a physical model for the system, starting with the simplest case of a uniform bulk sample.  Because the probe tip length is much less than the wavelength $\lambda \sim 4$ cm, we are able to perform a static calculation\cite{van_duzer} of the microwave electric fields.  Cylindrical symmetry further simplifies the problem to two dimensions.  Because of the complicated geometry of the probe tip in contact with a multi-layered sample, we use a finite element model on a grid.  We calculate the potential $\Phi$ in the region represented by the grid, taking into account any changes in the permittivity $\epsilon_{r}$, by solving Poisson's equation:
\begin{equation}
\nabla ^{2} \Phi + \frac{1}{\epsilon_{r}} (\nabla \Phi) \cdot (\nabla \epsilon_{r}) = 0 .
\label{poisson_eq}
\end{equation}
Using a rectangular grid, we represent the probe tip as a cone with a blunt end of radius $r_0$ (see Fig.\ \ref{model_diagram}), and perform the calculation in a spreadsheet program.  From Eq.\ (\ref{poisson_eq}), if $\Phi_{ij}$ is the potential at the cell at column $i$ and row $j$, $\Phi_{ij}$ is a weighted average of the values of the four adjacent cells:
\begin{equation}
\Phi_{ij} = \frac{\Phi_{i+1,j}(1+\frac{\Delta r}{r}) + \Phi_{i-1,j} + \Phi_{i,j-1} + \Phi_{i,j+1}}
{4 + \frac{\Delta r}{r}},
\label{cell_formula}
\end{equation}
where $r$ is the radius from the cylindrical coordinate axis, $\Delta r$ and $\Delta z$ are the spacing between cells in the $r$ and $z$ directions, respectively, and we take $\Delta r = \Delta z$.  Equation (\ref{cell_formula}) is the simplest case; in practice, the equation is complicated by interfaces between the dielectric sample ($\epsilon_{r} > 1$) and the air ($\epsilon_{r} = 1$), for example.  Near the probe tip (inside the dashed box in Fig.\ \ref{model_diagram}) where the electric field is the strongest, and hence, the most critical, the grid spacings $\Delta r$ and $\Delta z$ are small and uniform.  Inside this box we fix $\Delta z = \Delta z_{in} = 0.1 \mu$m, while the value of $\Delta r = \Delta r_{in}$ is uniform, but variable (0.5 $\mu$m$\ \lesssim \Delta r_{in} \lesssim 1.0 \mu$m), to allow for probes with different sharpnesses.  We represent the probe sharpness with the aspect ratio parameter $\alpha \equiv \Delta z_{in} / \Delta r_{in}$.

It is important for the boundaries of the grid to be sufficiently far away in order to minimize the effect of the chosen boundary conditions on the electric field near the probe tip.  To accomplish this, outside the dashed box in Fig.\ \ref{model_diagram}, the values of $\Delta r$ and $\Delta z$ continuously increase with distance away from the probe tip, allowing the outer radius of the grid to be at least 4 mm, and the height of the grid to be 2 mm.  The resulting grid consists of 84 $\times$ 117 cells, which is small enough to be a manageable calculation with a modern personal computer.  The top and outer boundary conditions were $d\Phi / dn = 0$ where $n$ is the coordinate normal to the edge.  At the bottom of the grid, which represents the bottom side of the 500 $\mu$m thick sample, we used $\Phi = 0$ for the boundary condition.  To match this condition, we always place the sample on top of a metallic layer for scanning; this has the added benefit of shielding the microscope from the effects of whatever is beneath the sample, which  could be difficult to model.

Two possible fitting parameters for the model are the aspect ratio $\alpha$ of the conical tip, and the radius $r_{0}$ of the blunt probe end.  We obtain a satisfactory fit with our data (see below) by fixing $r_{0} = (0.6\ \mu$m$) / \alpha$.  This leaves $\alpha$ as a single fitting parameter to represent all probes; typically we find that $1 < \alpha < 2$ for all of our probes.

Shown in Fig.\ \ref{field_image}(a) is the calculated electric field near the probe tip for a probe with $\alpha = 1$, and a sample with $\epsilon_{r} = 2.1$.  We notice that the fields are concentrated near the probe tip, as expected.  The spatial resolution is related to the size of the probe tip, as shown by this concentration of the fields near the tip.  Figure \ref{field_image}(b) and (c) show the calculated electric field as a function of radius away from the center of the probe, and depth in the sample, respectively, for two samples with $\epsilon_{r} = 2.1$ and 305.  For higher $\epsilon_{r}$, the fields are more highly concentrated, and fall off more quickly away from the probe, indicating that the spatial resolution of the microscope is higher for high $\epsilon_{r}$ materials.  From Fig.\ \ref{field_image}(b) and (c), the depth resolution appears to be slightly better than the lateral resolution.

\subsection{Calculating the Frequency Shift of the Microscope}

Using perturbation theory,\cite{Altshuler} we calculate the frequency shift of the microscope as a function of the fields near the probe tip.  We define $\epsilon_{r1}$ and $\epsilon_{r2}$ as the permittivities of two samples, the subscripts 1 and 2 indicating the unperturbed and perturbed system.  If $\bf E$$_{1}$ and $\bf E$$_{2}$ are the calculated electric fields inside the two samples, the frequency shift of the microscope upon going from sample 1 to sample 2 is\cite{Altshuler}
\begin{equation}
\frac{\Delta f}{f} \approx \frac {\epsilon _{0} }{4W} 
\int _{V _{S}} (\epsilon _{r2} - \epsilon _{r1} ) \textbf{E} _{1} \cdot \textbf{E} _{2} dV .
\label{delta_f_equation}
\end{equation}
where $W$ is the energy stored in the resonator, and the integral is over the volume $V_S$ of the sample. We calculate an approximate $W$ using the equation for the loaded Q of the resonator, $Q_{L} = \omega_{0} W/P_l$, where $\omega_{0}$ is the resonant frequency, and $P_l$ is the power loss in the resonator.  In our case, we use a bare 500 $\mu$m thick LaAlO$_{3}$ (LAO) substrate for the unperturbed system ($\epsilon_{r1} = 24$, $E_1$),\cite{Zuccaro} because its properties are well-characterized and it is a common substrate for oxide dielectric thin films.  

In the case of a bulk dielectric sample, we calculate the frequency shift $\Delta f$ upon replacement of the LAO substrate by the sample of interest with permittivity $\epsilon_{r2}$.  In the case of thin films on a LAO substrate, we calculate the frequency shift $\Delta f$ associated with replacing a thin top layer of the LAO substrate with a thin film of permittivity $\epsilon_{r2}$.

\section{Measuring the Permittivity of Bulk Samples}

To establish the validity of our microscope and model for performing permittivity measurements, we started with bulk dielectric materials. In order to obtain a comparison between the model presented above and experimental results, we scanned a series of 500 $\mu$m thick bulk dielectrics with known microwave permittivities:\cite{Konaka} Teflon ($\epsilon_r = 2.1$), MgO ($\epsilon_r = 10$), LAO ($\epsilon_r = 24$), and SrTiO$_3$ ($\epsilon_r = 300$).  In Fig.\ \ref{bulk_graph}, the experimental frequency shift data points are shown for three different probe tips with different aspect ratios $\alpha$.  Model fits are also shown, where only the value of $\alpha$ has been varied.  Note that the zero of frequency shift is for LAO ($\epsilon_r = 24$), the chosen unperturbed state of our resonator.

To measure the permittivity $\epsilon_{r}$ of a dielectric sample, we must first calibrate the probe (i.\ e.\ determine the parameter $\alpha$) using at least two samples with known $\epsilon_{r}$ and the same thickness.  Once $\alpha$ has been determined from this calibration, we can measure $\epsilon_{r}$ of any sample with any thickness, using the finite element model and Eq.\ (\ref{delta_f_equation}) to convert the measured $\Delta f$ to $\epsilon_{r}$.

We tested this method by measuring $\epsilon_r$ of a 500 $\mu$m thick KTaO$_3$ crystal, which is paraelectric with a cubic perovskite structure at room temperature ($T_c = 13$ K).\cite{Jona}  We found the permittivity to be $\epsilon_r = 262 \pm 20$, in agreement with microwave data in the literature ($\epsilon_r = 240$ at 9.4 GHz)\cite{Rupprecht} and low-frequency data obtained with a parallel-plate capacitor ($\epsilon_r = 260$ at 100 Hz, and $\epsilon_r = 238.5$ at 100 kHz).

\section{Imaging Permittivity of Thin Films}

For imaging thin films, we extended the finite element model to include a thin film on top of a dielectric substrate, as shown in Fig.\ \ref{thin_film_fig}.  For the purpose of thin film imaging, we consider the unperturbed system to be the probe tip in contact with a bare 500 $\mu$m thick LAO substrate; the perturbed system, in addition, has a dielectric thin film on top of the 500 $\mu$m thick LAO substrate.  Because the change in total thickness with the addition of the thin film is negligible compared to the 500 $\mu$m thick substrate, we can treat the thin film as the only perturbation to the system.  This perturbation due to the thin film is evident in Fig.\ \ref{thin_film_fig}, through the bending of the equipotential lines at the top and bottom surfaces of the thin film.  Thus, the volume $V_S$ in Eq.\ \ref{delta_f_equation} includes only the volume of the thin film, and we calculate the frequency shift $\Delta f$ associated with replacing a thin top layer of the LAO substrate with a thin film of permittivity $\epsilon_{r2}$.

The process for quantitative imaging of thin films is as follows.  First, we determine the probe's $\alpha$ parameter using the method described above using bulk samples.  Then, using the calculated electric fields from the finite element model for thin film samples, we use Eq.\ (\ref{delta_f_equation}) to calculate $\Delta f$, in the same way that we did for bulk samples, except that we integrate only over the volume of the thin film.  Using this model calculation, we obtain a functional relationship between $\Delta f$ and $\epsilon_{r2}$ of the thin film, much like the results shown in Fig.\ \ref{bulk_graph}.  This relationship is used to convert $\Delta f$ to $\epsilon_{r2}$, the permittivity of the thin film, after scanning the sample.

To test the thin film model, we imaged thin-film samples of SrTiO$_{3}$ and Ba$_{0.6}$Sr$_{0.4}$TiO$_{3}$ (BST) on LAO substrates.  The results are summarized in Table \ref{sto_bst_table}.  The thin-film permittivity was also measured at 10 kHz using Au interdigital electrodes deposited on the films.  Both our microwave measurements and the interdigital electrode measurements are primarily sensitive to the in-plane component of $\epsilon_r$ for these samples (see Sec.\ \ref{dir_sens_section}).  For the STO sample, the microwave permittivity is comparable to the low-frequency permittivity, showing that there is very little dispersion in this film.  The BST films both show significant dispersion, which is nonetheless within the range observed in the literature for similar films.\cite{Hoerman,Streiffer}

\section{Other Microscope Issues}
\subsection{Spatial Resolution
\label{spatial_res_section}}

Using the finite element model, we can calculate the spatial resolution of the microscope.  From Eq.\ (\ref{delta_f_equation}) we see that the important quantity is the electric field dot product contained in the integral.  Near the probe tip, the electric field is the strongest, and falls off nearly to zero at the outer boundary of the model grid.  Thus, most of the contribution to the integral will come from the region in the sample near the probe tip.  For our purposes, we define the lateral spatial resolution to be $2 r_{res}$, where the integral over the volume $V = \pi r_{res}^{2} d$ under the probe tip is equal to half of the integral over the volume of the whole sample.  For thin films, the depth $d$ is taken to be the thickness of the film, because we assume the substrate to be uniform and therefore not to contribute any features to the dielectric image.  For bulk samples, we choose $d = 2 r_{res}$, since we expect the vertical spatial resolution to be approximately equal to the lateral spatial resolution (see Fig.\ \ref{field_image}).

Figure \ref{spatial_res} shows the calculated spatial resolution for both a 500 $\mu$m thick bulk dielectric sample, and for a 400 nm thin film on a 500 $\mu$m thick LAO substrate, for a typical probe with $r_0 = 0.4\ \mu$m and an aspect ratio $\alpha = 1.5$.  We notice that the spatial resolution is the highest ($2 r_{res} \sim 1$ to 1.3 $\mu$m) for thin films; this value for $2 r_{res}$ agrees with experimental results, and is illustrated graphically in Fig.\ \ref{thin_film_fig} as a concentration of the electric field near the probe tip.  At high permittivities, the spatial resolutions for bulk and thin film samples converge to about 1.5 $\mu$m, which is approximately twice the tip bluntness of $2 r_0 = 0.8\ \mu$m.  Model results show that the spatial resolution for any $\epsilon_r$ is approximately proportional to $r_0$.

\subsection{Permittivity Tensor Directional Sensitivity
\label{dir_sens_section}}

Again using the model, we can determine the directional sensitivity of the microscope to the permittivity tensor by finding the relative magnitude of the radial ($I_{r}$) and vertical ($I_{z}$) components of the integral in Eq.\ (\ref{delta_f_equation}).  This ratio is equal to
\begin{equation}
\frac{I_r}{I_z} =
\frac{\int _{V _{S}} E_{r1} \cdot E_{r2} dV}
{\int _{V _{S}} E_{z1} \cdot E_{z2} dV} ,
\label{Ir_Iz_equation}
\end{equation}
where $E_{r1}$, $E_{r2}$, $E_{z1}$, and $E_{z2}$ are the unperturbed and perturbed radial and vertical electric fields, respectively.  This quantity $I_{r} / I_{z}$ is shown in Fig.\ \ref{directional_sens} as a function of sample permittivity for a 500 $\mu$m thick bulk dielectric and a 400 nm thin film on a LAO substrate.  For low-permittivity bulk samples, the normal ($z$) component of the permittivity tensor dominates; for most thin films, the in-plane (radial) component dominates, as shown in Fig.\ \ref{thin_film_fig}, where the electric field in the thin film is mainly in the horizontal direction.

We tested these model results by imaging a $5 \times 5 \times 10$ mm 0.1\% Ce-doped (Sr$_{x}$Ba$_{1-x}$)$_{1-y}$(Nb$_{2}$O$_{5}$)$_{y}$ (SBN) single crystal, with x = 0.61 and y = 0.4993.  This anisotropic crystal has specified permittivities of $\epsilon_{11} = 450$ and $\epsilon_{33} = 880$.\cite{SBN_info}  We found that the microscope frequency shift was $\sim 5\%$ larger (more negative) when the probe was in contact with the face with $\epsilon_{33}$ in-plane, relative to when the probe was in contact with the face with $\epsilon_{11}$ in-plane; this agrees well with our model results, which predict a difference of 7\%.

\subsection{Spherical Approximation for the Probe Tip}

As an alternative to our finite element model, an analytic solution for the electric field near the probe tip can be found by approximating the probe tip as a sphere of radius $r_{sph}$, and assuming an infinite sample.\cite{Gao,Gao2}  We tested the usefulness of a spherical approximation for the probe tip by replacing the blunt cone with a sphere in our finite element model.  To obtain the correct dependence of frequency shift on sample permittivity, it was necessary to include a second fitting parameter, the radius $r_{0}$ of a flat area of the sphere in contact with the sample.  Figure \ref{sphere_model}(a) shows the best fit of $\Delta f$ vs.\ $\epsilon_r$ for one of our probes for 500 $\mu$m thick bulk samples using a spherical tip model with $r_{sph} = 14\ \mu$m and $r_{0} = 0.4\ \mu$m.

The required radius of the sphere $r_{sph}$ is surprisingly large, so we calculated the spatial resolution of such a probe using the method described above, for a 400 nm thin film with $\epsilon_r = 300$ on a LAO substrate.  Shown in Fig.\ \ref{sphere_model}(b) is the integral inside a cylinder of radius $r_{res}$ as a fraction of the total integral in Eq.\ \ref{delta_f_equation}, for a spherical tip and a conical tip.  For the sphere-shaped probe, the spatial resolution was $2 r_{res} = 3\ \mu$m.  The conical probe model giving the best fit of $\Delta f$ vs.\ $\epsilon_r$ has $\alpha = 1.3$, giving a spatial resolution of 1.4 $\mu$m.  In addition, because of the slow fall-off of the integral as a function of radius for a spherical tip [Fig.\ \ref{sphere_model}(b)], imaging with such a tip would result in the smallest image features being smeared out to a diameter $>$ 10 $\mu$m.  Since we do not observe this effect, and find our spatial resolution to be $\lesssim$ 1.4 $\mu$m, we conclude that a cone is a much better approximation for the probe tip than a sphere, at least for the geometry of our system.  We believe that this is due to the long-range nature of electromagnetism, where the potential due to a point charge diminishes at the rate $1/r$, causing the sides of the cone-shaped tip to have an important contribution to the frequency shift.\cite{Lanyi}  Figure \ref{field_image}(a) illustrates this effect, by showing that even at a distance of 6 $\mu$m from the axis, the electric field is relatively strong at the cone surface.  

Also, a problem with an analytic solution using a sphere for the tip is the limitation to a single fitting parameter, the sphere radius $r_{sph}$.  For our probe tips, it was not possible to obtain a reasonable fit with the data on a $\Delta f$ vs.\ $\epsilon_r$ curve without an additional fitting parameter, such as the radius of a flat area on the sphere (a geometry which cannot be solved analytically).

The assumption of an infinite sample\cite{Gao,Gao2} can also pose problems.  We find that this is an unrealistic assumption, because the properties of whatever is directly beneath the sample substrate can have a measurable effect on the frequency shift of the microscope, which will affect quantitative imaging.  A better approach is to have a bulk conducting ground plane beneath the sample in the experiment and also in the model, so that everything beneath this ground plane can be safely ignored.

\section{Nonlinear Dielectric Imaging}
\label{nonlinear_imaging_section}

Electric field-dependent imaging can be accomplished by applying a voltage bias ($V_b$) to the probe tip via a bias tee in the resonator (see inset to Fig.\ \ref{schematic}).  A metallic layer beneath the thin film acts as a grounded counterelectrode.\cite{Cho}  In order to prevent the counterelectrode from dominating the microwave measurement (modeling of the system has shown this to be a potential problem\cite{Steinhauer3}), we use a high-sheet-resistance counterelectrode, making it virtually invisible to the microwave fields and causing minimal frequency shift.  This is demonstrated in Fig.\ 2(a) in Ref.\ \onlinecite{Steinhauer1}, where the frequency shift in the high-sheet-resistance limit is shown to saturate at a value dependent only on the substrate.  As a result, the presence of the thin-film counterelectrode can be safely ignored in the finite element model described above.  Because the counterelectrode is immediately beneath the dielectric thin film, the applied electric field is primarily in the vertical direction, unlike the microwave electric field, which is mainly in the horizontal direction for thin films with large permittivies (Fig.\ \ref{directional_sens}).  Also, by simulating the applied field using a finite element model similar to that presented above, we find that the applied electric field beneath the probe tip is approximately uniform and equal to $E_b = V_b / t_f$, where $t_f$ is the thickness of the dielectric thin film.

By modulating the bias voltage applied to the probe tip, we can extract nonlinear terms in the permittivity.  Expanding the electric displacement {\bf D} in powers of the electric field {\bf E}, and keeping only the nonzero terms,\cite{Cho} we have
\begin{equation}
D_1 ({\bf E}) = \epsilon _{11} E_1 + \frac{1}{2} \epsilon _{113} E_1 E_3 +
\frac{1}{6} \epsilon _{1133} E_1 E_3^2 + ...
\label{D_expansion_equation}
\end{equation}
where $E_1$ is the rf electric field in the $r$ direction, and $E_3 = E_b$ is the applied bias electric field in the $z$ direction.

Adding a low-frequency oscillatory component ($\omega_{b} =$ 1 kHz, with an amplitude of $\tilde{V}_{b} = 1$ V, in our case) to the bias voltage, the applied electric field is $E_{b} = E_{b}^{dc} + \tilde{E}_{b} \cos \omega_{b} t$.  We find that the effective rf permittivity is then\cite{Cho}
\begin{eqnarray}
\epsilon_{rf} = &&\epsilon_{11} + \frac{1}{2} \epsilon_{113} E_{b}^{dc} + \epsilon_{1133} \left ( \frac{(E_{b}^{dc})^2}{6} + \frac{(\tilde{E}_{b})^2}{12} \right ) \nonumber\\
&&+ (\frac{\epsilon_{113}}{2} + \frac{\epsilon_{1133} E_{b}^{dc}}{3}) \tilde{E}_{b} \cos(\omega_{b} t)\nonumber\\
&&+ \frac{1}{12} \epsilon_{1133} \tilde{E}_{b}^2 \cos(2 \omega_{b} t) + ...
\label{eps_rf_equation}
\end{eqnarray}
We note that the components of $\epsilon_{rf}$ at $\omega_b$ and $2 \omega_b$ are approximately proportional to $\epsilon_{113}$ and $\epsilon_{1133}$, respectively.  Expanding the resonant frequency of the microscope as a Taylor series about $f_0(\epsilon_{rf} = \epsilon_{11})$, we have
\begin{equation}
f_0[\epsilon_{rf}(t)] = f_0(\epsilon_{11}) + \left. \frac{d f_0}{d\epsilon_{rf}}\right |_{\epsilon_{rf} = \epsilon_{11}} [ \epsilon_{rf}(t) - \epsilon_{11} ] + ...
\label{f0_equation}
\end{equation}
Substituting Eq.\ \ref{eps_rf_equation} into Eq.\ \ref{f0_equation}, and keeping only the larger terms, 
\begin{eqnarray}
f_0(t) \approx && {\rm constant} + \frac{1}{2}\epsilon_{113} \tilde{E}_{b} \left. \frac{df_0}{d\epsilon_{rf}} \right |_{\epsilon_{11}} \cos(\omega_b t) \nonumber\\
&& + \frac{1}{12} \epsilon_{1133} \tilde{E}_{b}^2 \left. \frac{df_0}{d\epsilon_{rf}} \right |_{\epsilon_{11}} \cos(2 \omega_b t) .
\end{eqnarray}
Thus, the components of the frequency shift signal at $\omega_b$ and $2 \omega_b$ can be extracted to determine the nonlinear permittivity terms $\epsilon_{113}$ and $\epsilon_{1133}$.  These nonlinear terms can be measured simultaneously with the linear permittivity ($\epsilon_{11}$) while scanning.  

As an alternative, the electric field {\bf E}$_b$ could be applied in the horizontal direction using thin film electodes deposited on top of the dielectric thin film.  The advantage in this case is that diagonal nonlinear permittivity tensor terms could be measured, such as $\epsilon_{111}$ and $\epsilon_{1111}$;  the disadvantage is that imaging is limited to the small gap between the electrodes.

\section{An Application:  Measuring Recovery of Dielectric Properties in a Thin Film After Annealing}

To demonstrate the usefulness of the microwave microscope, we scanned a sample consisting of a laser-ablated 370 nm thick Ba$_{0.6}$Sr$_{0.4}$TiO$_3$ (BST) thin film on a 70 nm La$_{0.95}$Sr$_{0.05}$CoO$_3$ (LSCO) counterelectrode.  The substrate is LaAlO$_3$ (LAO).  The films were deposited at 700 $^{\circ}$C in 200 mTorr of O$_2$.  The sheet resistance of the LSCO counterelectrode is about 400 $\Omega/\Box$, sufficiently large to render it invisible at microwave frequencies.

Figure \ref{bst_images}(a) shows a schematic diagram of a $76.5 \times 20\ \mu$m$^2$ region of the thin film sample which we scanned.  The gray-shaded areas indicate regions which were milled through the BST layer using a gallium focused ion beam (FIB).  There is a 1 $\mu$m wide line and a corner of a 5 $\mu$m wide ``frame" surrounding an untouched $20 \times 20\ \mu$m$^2$ region.  Fig.\ \ref{bst_images}(b) shows a permittivity image of the region sketched in (a).  The narrow line shows up as a wider band with low permittivity, a sign that the region near the line was damaged by the gallium ion beam tails\cite{Ward} during milling.  The wide frame appears as a double line in (b) due to edge effects at the edge of the milled area; also, the value of $\epsilon_r$ shown in this wide frame region is invalid, since there is no dielectric thin film there.  Several low-permittivity regions appear randomly scattered over the scan area.  We have shown previously\cite{Steinhauer3} that these are particles (``laser particles") on the surface which probably accrued during pulsed laser deposition. 

After acquiring the image shown in Fig.\ \ref{bst_images}(b), we annealed the sample at 650 $^{\circ}$C in air for 20 minutes.  Afterward, we scanned the same region again; the resulting image is shown in (c).  The overall thin film permittivity has increased between (b) and (c), and the narrow line has become less prominent, indicating that perhaps the permittivity of the damaged region was partially restored.  

Figure \ref{bst_images}(d) and (e) show nonlinear dielectric images ($\epsilon_{113}$) taken with an applied dc bias of -3.5 V, giving an average vertical dc electric field of -95 kV/cm beneath the probe tip.  Figure \ref{bst_images}(d) was acquired before the sample was annealed, and shows that the tunability has been destroyed near the narrow FIB milled line, and within about 12 $\mu$m of the FIB milled frame.  Figure \ref{bst_images}(e), which was acquired after annealing, shows that tunability has been restored in the damaged regions, and slightly improved in the undamaged regions.

Figure \ref{bst_graphs} shows hysteresis loops taken after the sample was annealed, at the point marked by a ``+" in Fig.\ \ref{bst_images}(c) and (e).  The thin film permittivity is shown as a function of applied electric field in Fig.\ \ref{bst_graphs}(a).  As expected, the permittivity decreases when an electric field is applied.  The small amount of hysteresis is probably due to the broadened transition temperature of the nominally paraelectric thin film.  The first nonlinear dielectric term $\epsilon_{113}$ is shown in Fig.\ \ref{bst_graphs}(b), while the second nonlinear term $\epsilon_{1133}$ is shown in (c).  We notice that $\epsilon_{113}$ is close to zero for zero applied field, which would be expected for a material with no spontaneous polarization.\cite{Cho}  In addition, we expect the sign of $\epsilon_{113}$ to be sensitive to the direction of the polarization, which is evident in (b).  The second nonlinear term, on the other hand, is nonzero at zero applied field.  The curves in Fig.\ \ref{bst_graphs} are centered at a nonzero field $\sim -20$ kV/cm, probably because the asymmetric capacitor electrodes induce unequal charges at the two electrodes.\cite{Miranda,Steinhauer3}  The observed tunability of $\epsilon_r$ is small ($\sim\ 2\%$) probably because we are measuring an off-diagonal nonlinear component of the permittivity tensor ($\epsilon_{113}$).\cite{Steinhauer3}

We have evidence that the nonlinearity signal at $\omega_b$ (see Sec.\ \ref{nonlinear_imaging_section}) is related to the normal component of ferroelectric polarization, allowing us to image domain structures in bulk crystals.

\section{Sensitivity of the Microscope}

We find the sensitivity of the microwave microscope by observing the noise in the dielectric permittivity and tunability data.  For a 370 nm thin film on a 500 $\mu$m thick LAO substrate, with an averaging time of 40 ms, we find that the permittivity sensitivity is $\Delta\epsilon_{r} = 2$ at $\epsilon_{r} = 500$, and the nonlinearity sensitivity is $\Delta \epsilon_{113} = 10^{-3}$ (kV/cm)$^{-1}$.

\section{Future Improvements}

The sensitivity of the microscope to sample permittivity could be improved by decreasing noise and drift in the electronics, using a high-stability microwave source, for example.  A piezoelectric scanning system will reduce vibration noise and mechanical drift caused by stepper motors.  The spatial resolution is currently limited by the probe tip radius, which is made worse by contacting the sample;  it could be improved by operating the microscope out of contact with a distance feedback mechanism.  The latter two improvements could reduce the line-to-line registration drift seen in Fig.\ \ref{bst_images}.

\section{Acknowledgments}

The authors would like to thank Hans Christen of Neocera, Inc., for his low-frequency measurements on the KTaO$_3$ crystal.  This work has been supported by NSF-MRSEC grant No.\ DMR-9632521, NSF grants No.\ ECS-9632811 and DMR-9624021, and by the Maryland Center for Superconductivity Research.

\begin{table}
\caption{Comparison of microwave permittivity $\epsilon_r$ of thin-films on LaAlO$_3$, measured with the microwave microscope, and low-frequency permittivity, measured using interdigital electrodes.}
\label{sto_bst_table}
\begin{tabular}{llll}
Film material& Thickness (nm) & Microwave $\epsilon_r$ & $\epsilon_r$ at 10 kHz\\
\tableline
SrTiO$_{3}$                   & 440 & $145 \pm 26$ at 7.2 GHz & 150\\
Ba$_{0.6}$Sr$_{0.4}$TiO$_{3}$ & 300 & $388 \pm 14$ at 8.1 GHz & 700\\
Ba$_{0.6}$Sr$_{0.4}$TiO$_{3}$ & 400 & $573 \pm 27$ at 8.1 GHz & 1030\\
\end{tabular}
\end{table}

\begin{figure}[tbp]
\caption{Schematic of the microwave microscope.  The inset shows a close-up of the probe and a thin-film sample, along with the probe bias voltage, $V_{b}$.}
\label{schematic}
\end{figure}

\begin{figure}[tbp]
\caption{The geometry of the finite element relaxation model (not to scale).  The left border represents the cylindrical coordinate axis.  Inside the dashed box, the grid cell spacings, $\Delta r$ and $\Delta z$ are constant, while outside, they continuously increase toward the boundaries.}
\label{model_diagram}
\end{figure}

\begin{figure}[tbp]
\caption{(a) The calculated electric field magnitude near the probe tip (in the region indicated by the dashed box in Fig.\ \ref{model_diagram}) for a sample with permittivity $\epsilon_{r} = 2.1$, and a probe with aspect ratio $\alpha = 1$ and tip bluntness $r_0 = 0.6\ \mu$m.  The electric field magnitude as a function of radius (b) and depth in the sample (c) are shown for samples with $\epsilon_{r} = 2.1$ and 305.}
\label{field_image}
\end{figure}

\begin{figure}[tbp]
\caption{Model results (lines) vs. data (symbols) with 500 $\mu$m thick bulk dielectrics for three different probe tips with different values of the aspect ratio $\alpha$.  Frequency shifts are relative to a LaAlO$_3$ (LAO) sample, with $\epsilon_{r} = 24$.}
\label{bulk_graph}
\end{figure}

\begin{figure}[tbp]
\caption{The electric field near a probe tip in contact with a thin film with $\epsilon_r = 300$ on a substrate with $\epsilon_r = 24$.  The region shown is $6 \times 6.8\ \mu$m$^2$.  The equipotential lines are shown; electric field lines are perpendicular to these lines.}
\label{thin_film_fig}
\end{figure}

\begin{figure}[tbp]
\caption{Spatial resolution of the microscope ($2 r_{res}$) as a function of sample permittivity, for a probe tip with aspect ratio $\alpha = 1.5$ and radius $r_0 = 0.4\ \mu$m (see Fig.\ \ref{model_diagram}).  These model results are given for a 500 $\mu$m thick bulk sample, and a 400 nm thin film on top of a 500 $\mu$m bulk LAO substrate ($\epsilon_r = 24$).}
\label{spatial_res}
\end{figure}

\begin{figure}[tbp]
\caption{The directional sensitivity of the microscope to the sample's permittivity tensor.  The ratio $I_r/I_z$ gives the relative contribution of the radial (in-plane) to the vertical (normal) components of $\epsilon_r$ for a sample with isotropic $\epsilon_r$.  Results are given for a 500 $\mu$m thick bulk sample, and a 400 nm thin film on top of a 500 $\mu$m bulk LAO substrate.}
\label{directional_sens}
\end{figure}

\begin{figure}[tbp]
\caption{(a) Data vs.\ model results, using the approximation of a sphere for the probe tip, with 500 $\mu$m thick bulk dielectric samples.  The spherical tip parameters are the sphere radius $r_{sph} = 14\ \mu$m, and contact area radius $r_0 = 0.4\ \mu$m.  (b) The contribution to the electric field integral in Eq.\ \ref{delta_f_equation} inside a radius $r_{int}$ as a fraction of the total integral.  Results are given for the sphere-shaped probe tip, and a blunt cone-shaped tip (with $\alpha = 1.3$), for a 400 nm thin film of permittivity $\epsilon_r = 300$ on a 500 $\mu$m thick LAO substrate.  The spatial resolution for the spherical tip is $\sim 3\ \mu$m, while for the conical tip it is a more realistic $\sim 1.4\ \mu$m.}
\label{sphere_model}
\end{figure}

\begin{figure}[tbp]
\caption{Images of a 370 nm thick Ba$_{0.6}$Sr$_{0.4}$TiO$_3$ (BST) thin film on a 70 nm La$_{0.95}$Sr$_{0.05}$CoO$_3$ (LSCO) counterelectrode on a LaAlO$_3$ (LAO) substrate.  All images are of the same 76.5 $\times 20\ \mu$m$^2$ region. (a) Schematic diagram of the focused ion beam (FIB) milled regions in the scan region. (b) Scan of the thin film permittivity before annealing. (c) Thin film $\epsilon_r$ after annealing at 650 $^{\circ}$C in air for 20 minutes. (d) Thin film tunability as shown in the nonlinear dielectric term $\epsilon_{113}$, before annealing.  The low-tunability regions were damaged during FIB milling. (e) Nonlinear dielectric ($\epsilon_{113}$) image after annealing, showing that the tunability has been restored. }
\label{bst_images}
\end{figure}

\begin{figure}[tbp]
\caption{Hysteresis loops at the location marked with a ``+" in Fig.\ \ref{bst_images}.  Permittivity (a), first nonlinear permittivity term $\epsilon_{113}$ (b), and second nonlinear term $\epsilon_{1133}$ (c) of the thin film as a function of the applied electric field $E_b^{dc}$.}
\label{bst_graphs}
\end{figure}

\begin{center}
\leavevmode
\epsffile[58 93 550 720]{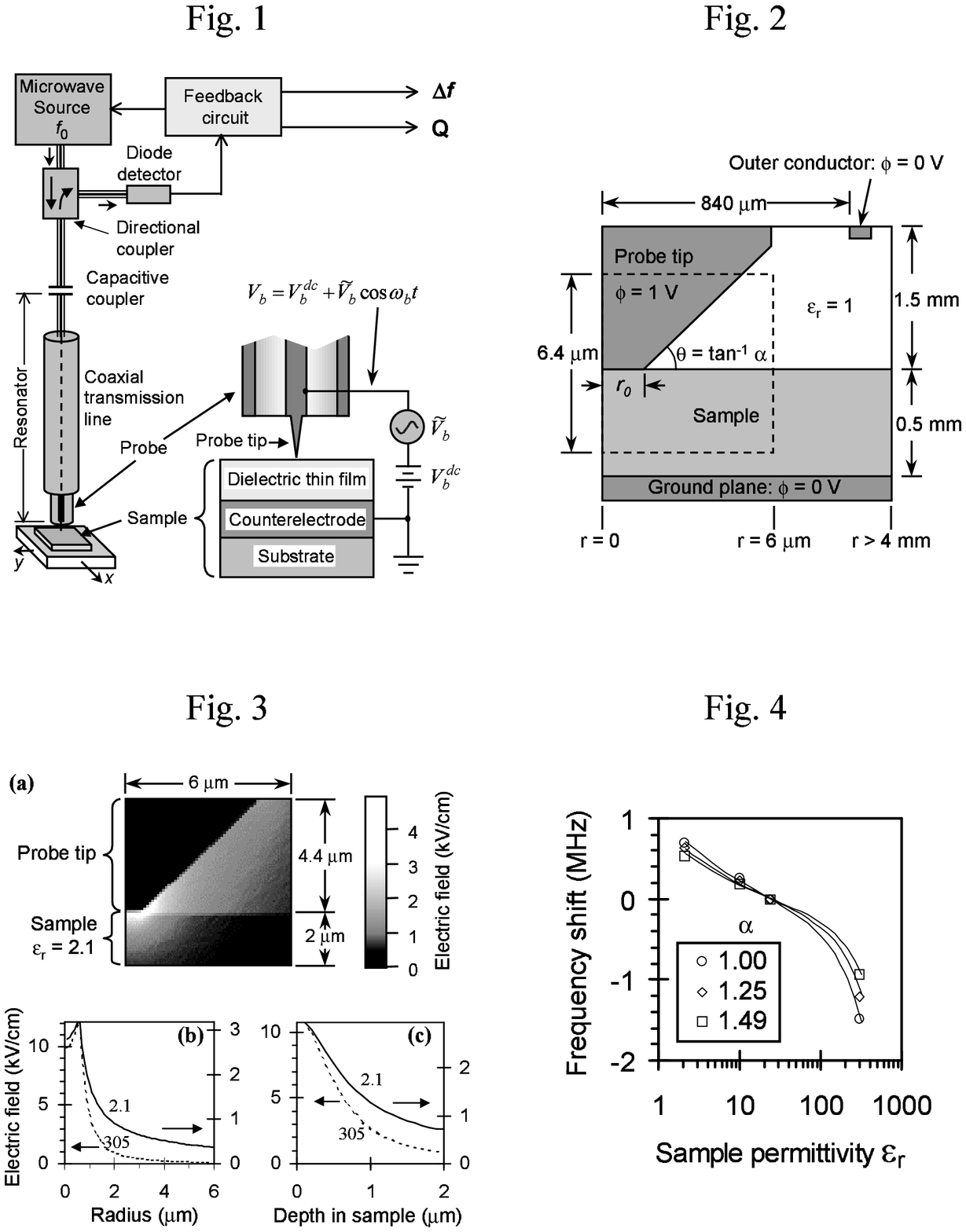}
\end{center}

\begin{center}
\leavevmode
\epsffile[62 162 535 686]{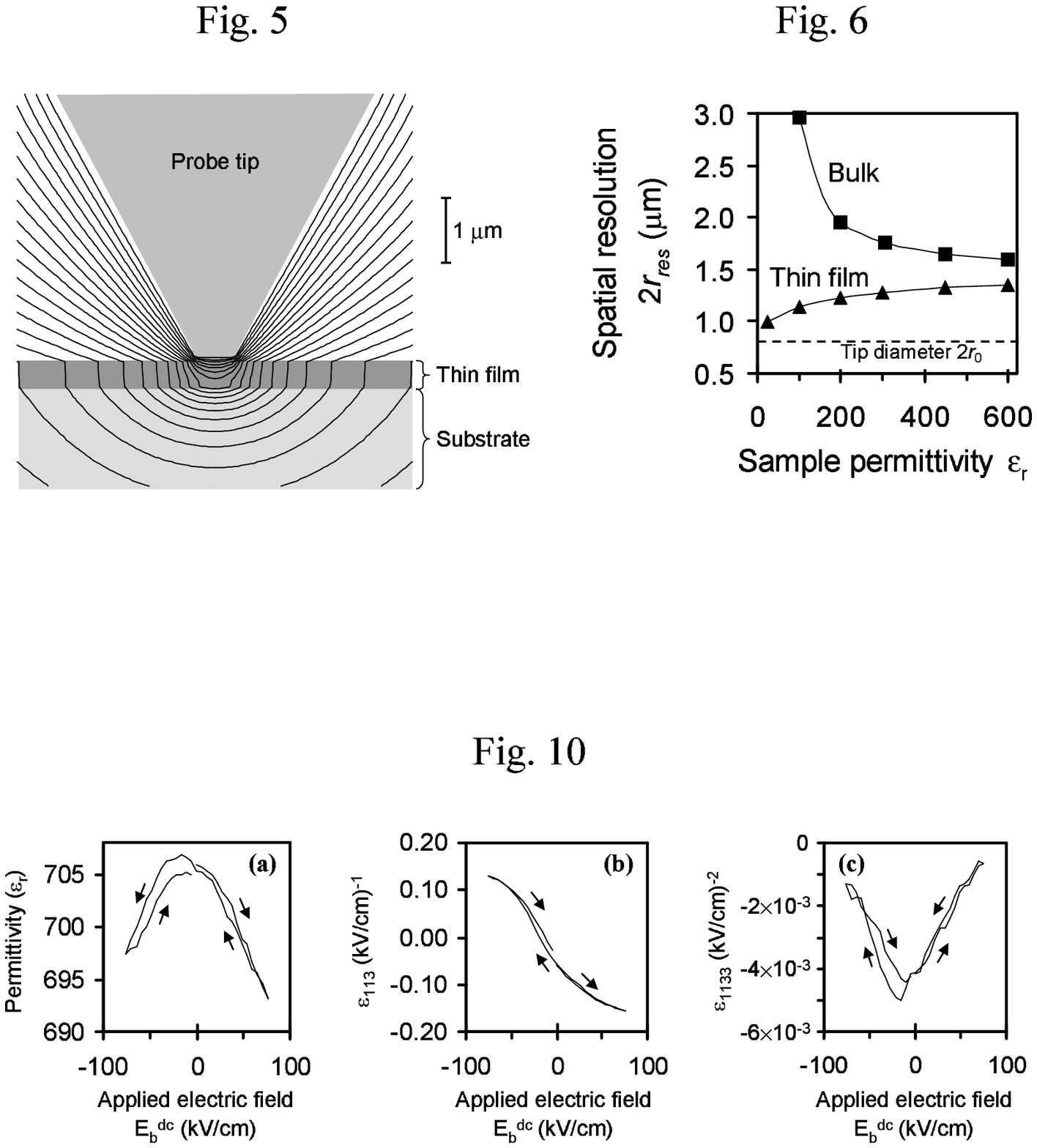}
\end{center}

\begin{center}
\leavevmode
\epsffile[53 254 564 702]{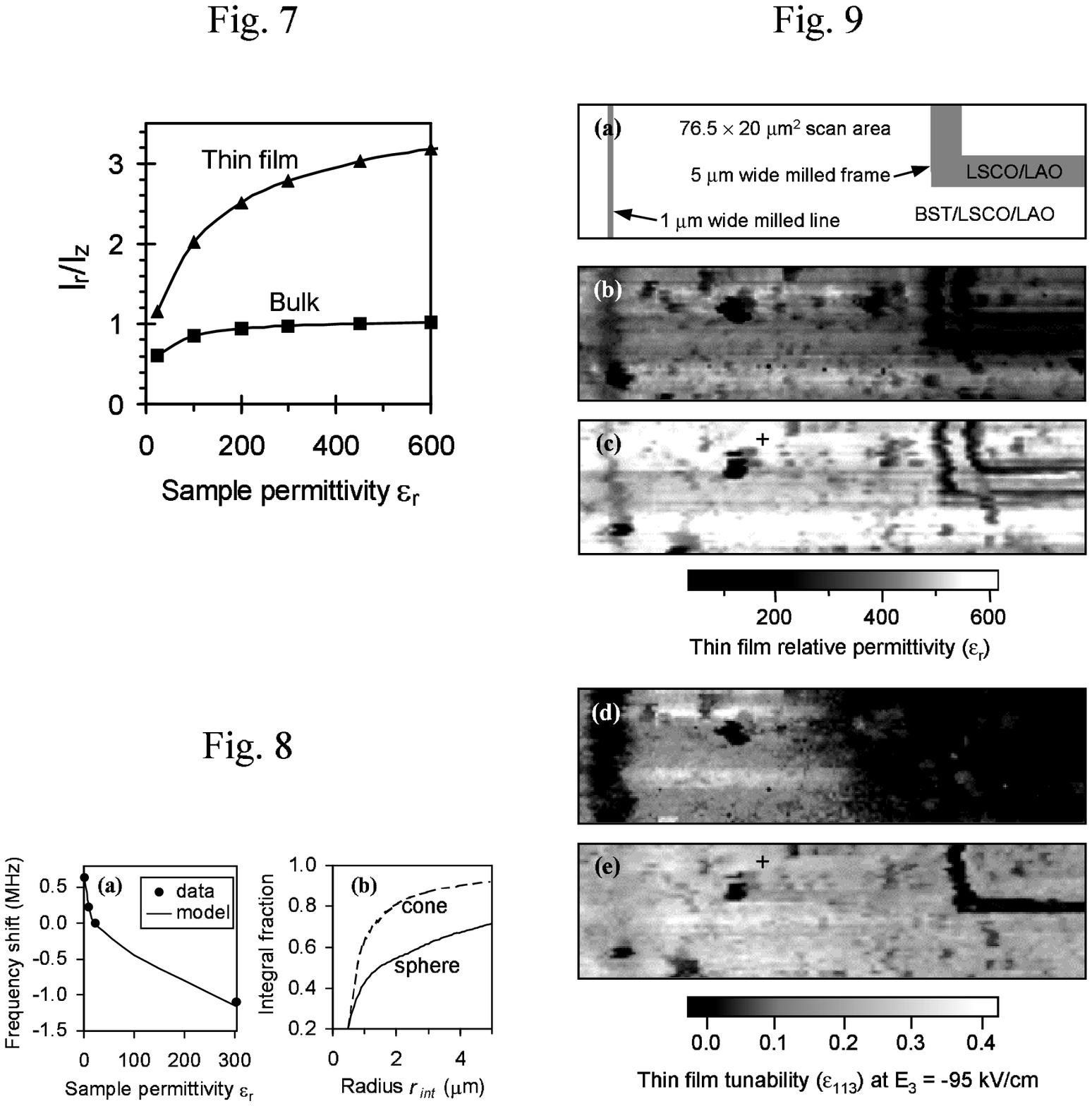}
\end{center}


\begin{references}
\bibitem{force_ref} Although the magnitude of the force between the sample and the probe tip affects the measurement of the permittivity, causing a small error, this error is comparable to or smaller than the system noise when the force is $\sim 75\ \mu$N or less.  In addition, we have found that a force of at least 50 $\mu$N is necessary to minimize the effects of vibrations in the environment of the microscope.
\end{references}
\end{document}